\documentclass[runningheads]{llncs}
\usepackage[T1]{fontenc}
\usepackage{graphicx}
\usepackage{cite}
\begin{document}
\title{Real-Time Cyberattack Detection with Offline and Online Learning\thanks{This research has been supported by the European Commission H2020 Program under the IoTAC Research and Innovation Action, under Grant Agreement No. 952684.}}
%
%
\author{Erol Gelenbe\inst{1,2,3}\orcidID{0000-0001-9688-2201} \and Mert Nak\i p\inst{1}\orcidID{0000-0002-6723-6494}}
\authorrunning{Erol Gelenbe and Mert Nak\i p}
%
\institute{Institute of Theoretical and Applied Informatics, Polish Academy of Sciences (PAN), ul. Baltycka 5, 44100 Gliwice, Poland \and
Lab. I3S, Universit\'{e} C\^{o}te d'Azur, Grand Château, 06103 Nice Cecex 2, France 
\and Department of Computer Engineering, Ya\c{s}ar University, Üniversite Cad. No: 37-39, A\~{g}a\c{c}l\i~Yol, Bornova, Izmir, Turkey\\ 
\email{seg@iitis.pl,mnakip@iitis.pl}}
\maketitle              
\begin{abstract}
This paper presents several novel algorithms for real-time cyberattack detection using the Auto-Associative Deep Random Neural Network, which were developed in the HORIZON 2020 IoTAC Project. Some of these algorithms  require offline learning, while others require the algorithm to learn during its normal operation while it is also testing the flow of incoming traffic to detect possible attacks. Most of the methods we present are designed to be used at a single node, while one specific method collects data from multiple network ports to detect and monitor the spread of a Botnet. The evaluation of the accuracy of all the methods is carried out with real attack traces. These novel methods are also compared with other state-of-the-art approaches, showing that they offer better or equal performance, at lower computational learning and shorter detection times as compared to the existing approaches. 
\end{abstract}
\section{Introduction}

As the application areas of the Internet expand, and the number of IoT devices increases rapidly,  52 \% of the devices deployed in the IoT devices are expected  \cite{Cisco2020} to be of low-cost and have low-maintenance, and typically perform a single task at a time.
This sector of the IoT is often known as the ``Massive IoT'' which is largely composed of devices which cannot run complex real-time attack detection or prevention algorithms. Thus the IoT and the Internet as a whole are becoming increasingly vulnerable to  cyberattacks \cite{Frustaci, Lin_IoTsurvey, Yang_Securitysurvey}.

Another industry report \cite{hp} states that 70 \% of IoT devices are vulnerable to attacks, such as common Denial of Service (DoS) or Distributed Denial of Service attacks (DDoS) \cite{Douligeris}. According to some sources these represent 20 \% of all attacks \cite{Benzarti}, in which attackers or malicious devices send numerous meaningless requests to prevent the targeted devices for arrying out their normally required activities. 
In addition, many of these attacks can  inject  malware \cite{CISA, Carl}, as with Botnets which cause the victim to become an attacker that in turn generates a flood of traffic against other IP addresses and servers. A ajor example is the 2016 massive Mirai Botnet attack targeted Domain Name System (DNS) provider Dyn \cite{Margolis}, compromising numerous websites and servers of leading companies \cite{Biggs,Hackett}. Botnets also have other undesirable effects, such as increasing the power consumption and memory occupancy of the devices \cite{Tushir_impactsOfMirai}. Thus in detecting and countering a cyberattack, it is important to identify not the malicious packets, and also the other compromised IP
addresses which themselves become attackers against IoT networks.

\subsection{Related Work}

Over recent years, much work has studied the characteristics of Botnet attacks \cite{Manos}, and analyzed the charateristics of attack traffic flows \cite{kumar2019}. The source code of such attacks was examined in  \cite{Sinanovic} while in \cite{Margolis} their capabilities and impact are studied. Some authors \cite{Ahmed2019} also suggest using blockchains to protect IoT devices. 

Other work has used different machine learning methods to detect Botnet attacks including  k-Nearest Neighbours (KNN), Support Vector Machines (SVM), Decision Trees (DT) and Multi-Layer Perceptrons (MLP). Thus the work in \cite{Tuan} compares the performance of classification models and Neural Networks (NNs), and in \cite{Letteri} NNs are used to to detect Mirai Botnet 
attacks against Software Defined Networks.  While the work in \cite{Sriram} uses a deep MLP, in \cite{Soe} NNs and Naive Bayesian networks (NB) were used with a sequential architecture, and in \cite{Tzagkarakis} Botnets are detected via a sparse representation framework. To detect Botnet attacks using deep learning, in \cite{McDermott} text recognition is carried out with a bidirectional Long-Short-Term Memory (LSTM). Furthermore in \cite{Liu_botnet} a Convolutional Neural Network (CNN) is combined with a feature transformation, while the work in \cite{Parra} cascaded a CNN with a LSTM. In \cite{Htwe} Classification and Regression Trees (CART) were used, while in \cite{Banerjee} DTs, Gradient Boosting and Random Forest were considered, and in \cite{Prokofiev} Logistic Regression was used. 

In \cite{Kumar_MiraiBots} an optimization-based method scans the number of destination ports in the headers of selected IoT packets to detect Bots infected by Mirai. Others \cite{Chatterjee_bot} developed a traffic analysis technique based on evidence theory in order to detect compromised devices selecting the rarest features from the set of traffic features including number of reconnections, transport layer protocol, and source/destination ports. In ~\cite{Nguyen_federated}, federated learning and language analysis is used to detect malicious devices using pre-identified individual device types, and in  \cite{Abhishek_gateway} compromised gateways that monitor the downlink channels in an IoT network are detected. In \cite{Taneja} compromised mobile devices are deected taking into account their location, classifying locations and their possible changes  as unusual behavior.

In earlier work, a Random Neural Network gradient descent learning technique \cite{RNN,RNN2} had been used to detect SYN type DoS attacks \cite{Spilios}. However the approaches discussed in this paper are based on a more elaborate machine learning model, the Auto-Associative Dense Random Neural Network (AADRNN), that was first developed in \cite{Brun} using a Deep and Dense Random Neural Network \cite{Deep1,Deep2}.
In addition, it uses  an original characterization of network traffic proposed in \cite{nakip_MIRAI} to detect various types of attacks.

\section{Contributions of the Present Paper}

This paper presents novel Attack Detection (AD) algorithms that were developed within the IoTAC Project funded by the Horizon2020 Programme,  based on Auto-Associative version 
of the Deep Random Neural Network (AADRNN).  The three sets of results we present 
show the ability of this AD learning approach to detect Botnet attacks with online learning,  as well as to simultaneously detect different types of attacks, and its ability to identify compromised IoT devices. 

The main advantage of this approach is the use of an auto-associative network (namely, the AADRNN), that only needs to learn from the  legitimate network traffic, without needing to learn from the different types of attacks that may occur, and then generalizes what it has learned only from nomal traffic to differentiate between benign and malicious packets so as to identify compromised devices. Thus: 
\begin{enumerate}
	\item Training of the proposed algrithms does not require attack traffic data, which saves considerably on the time needed to collect it to create it artificially, and also on the algorithms' learning times.
	\item It follows that a give AADRNN after training with normal data can accurately detect various types of attacks. 
	\item The proposed AD can be trained in parallel to its real-time operation (when it provides detection results) using the ongoing legitimate network traffic. 
\end{enumerate}

\section{IoTAC's Attack Detector Algorithms}

IoTAC's AD has been designed to recognize traffic patterns through specially defined metrics and is trained with normal traffic to create an Auto-Association Dense RNN (called AADRNN). Thus, this AD can recognize malicious traffic even if the characteristics of an attack are unknown and no pre-collected attack data is available.

\begin{figure}
\includegraphics[scale=0.2]{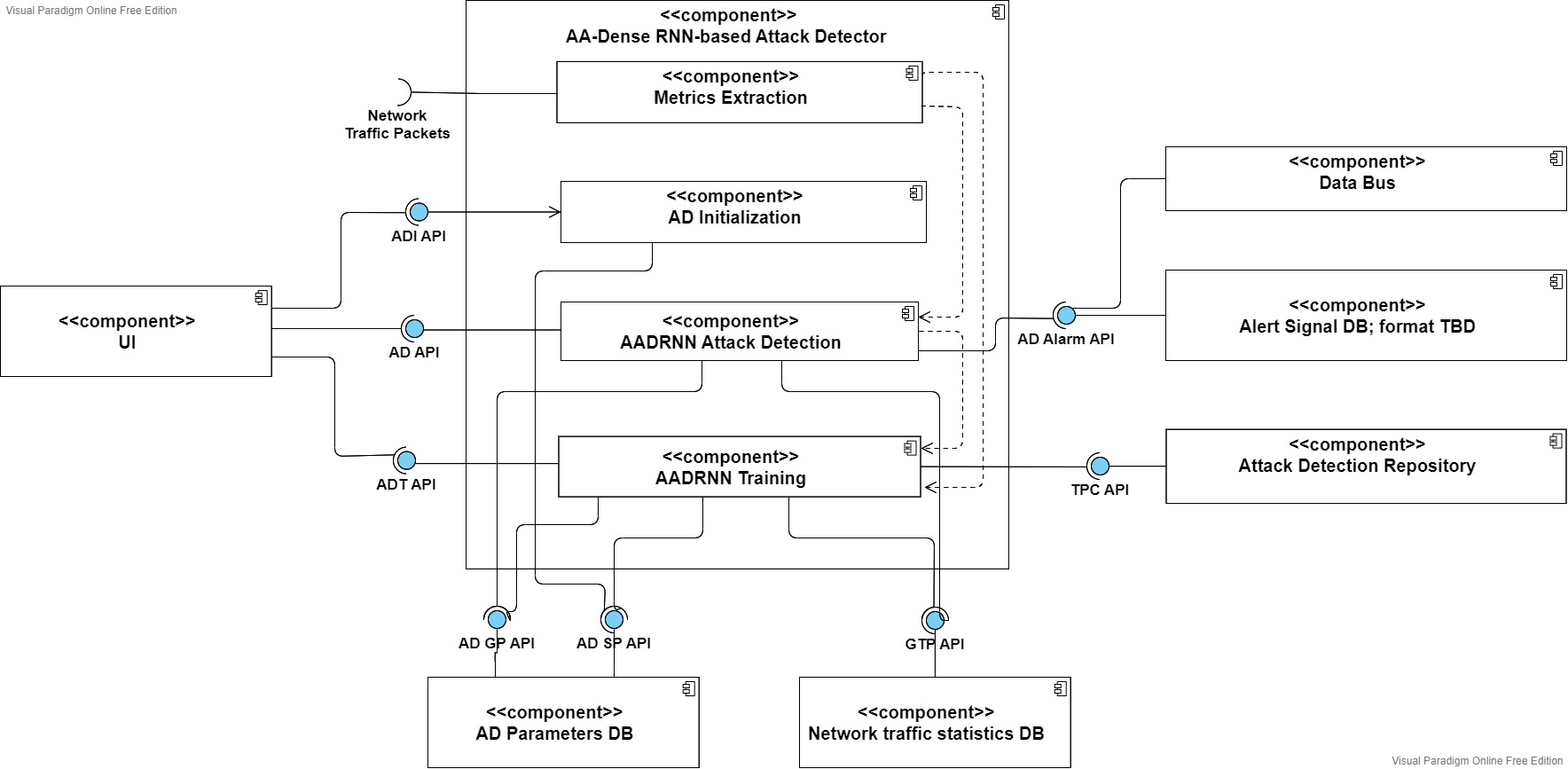}
\caption{Component diagram of AD} \label{fig:component_diagram}
\end{figure}

Figure~\ref{fig:component_diagram} displays the component diagram of AD including the subcomponents, APIs, external databases, and user interfaces. As shown in this figure, the AD component is comprised of four subcomponents: Metrics Extraction, AD Initialization, AADRNN Attack Detection, and AADRNN Training. 

\subsection{AD Initialization}

At the first use of the AD module, parameter initialization is required for the methods utilized in AD, such as Metric Extraction and AADRNN. To this end, the AD Initialization subcomponent is developed to set the parameters of AD as predefined values. It also calculates the initial values of scaling factors used to normalize the metric values through historical normal traffic for a fixed length time window.

\subsection{Metric Extraction}

First, the Metric Extraction subcomponent calculates three metrics proposed in \cite{nakip_MIRAI} to capture the footprints of the Mirai Botnet attack on the network traffic:
\begin{enumerate}
	\item The total size of the latest packets,
	\item The average inter-transmission times of the latest packets,
	\item The total number of packets that are transmitted in a constant-length time window.
\end{enumerate}
These metrics are specifically defined to represent network traffic in a way that the differences between attack and normal traffic become more visible while they can be calculated using only the header information of the traffic packets. Therefore, these metrics can be easily calculated without the need for any sensitive or device-specific information, thus preventing AD from making biased decisions, remaining anonymous regarding packet content and communicating devices, and suitable for real-time operation on lightweight systems.

\subsection{AADRNN Attack Detection}

The AADRNN Attack Detection subcomponent uses a trained AADRNN and a simple decision-making algorithm. Based on the extracted metrics, the AADRNN in this subcomponent predicts the expected metric values for the normal operation of the network. That is, AADRNN provides the metric values expected to be obtained if all packets are benign. The decision-making algorithm considers these expected metric values as a baseline for actual metric values expecting to observe a significant difference between them for malicious packet transmission. Accordingly, this simple algorithm calculates the weighted average of the absolute differences between the expected and actual metric values and applies a threshold to the mean.

\subsection{AADRNN Training}

The AADRNN model of AD is trained incrementally in parallel to the real-time operation of AD through ADT API using only normal traffic to learn its metrics. To this end, we have developed an incremental semi-supervised training procedure \cite{incremental2022} based on a reconstruction problem \cite{Deep1}. Specifically, our incremental training algorithm stores historical normal traffic for fixed-length time windows and updates the connection weights of the AADRNN for this traffic at the end of each window. During this incremental training, the connection weights of this model are updated to reconstruct the metrics of normal traffic from a noise-added version of these metrics. In this way, we have obtained an auto-associative network that is able to retrieve normal metrics from any metrics (which might be for either normal or malicious traffic) provided as input.
Using an auto-associative network with the proposed metrics provides the following benefits for the AD module:

\begin{enumerate}
	\item Since the defined traffic metrics are calculated using only the high-level (anonymous) packet information, the AD module does not use and does not require knowledge of network architecture or device specifications.
	
	\item The AD module has the ability to react to malicious acitivities only by learning the traffic patterns during the normal operation of the IoT network. Therefore, 
	\begin{enumerate}
		\item it does not require data on "attack traffic" for training,
		\item	it has no bias for any particular type of attack, and 
		\item	a single AD can detect several types of attacks simultaneously. 
	\end{enumerate}
\end{enumerate}

\section{AADRNN-based Algorithm Detects Several Types of Cyberattacks} 

The AADRNN-based design of AD provides an opportunity to adapt its usage for detecting different types of attacks in various environments. Thus we have extended the design of the AADRNN attack detector for usage in three areas: 
\begin{itemize}
	\item Botnet Attack Detection with Online Learning, 
	\item Simultaneous Detection of Different Attack Types, and 
	\item Compromised IoT Device Identification. 
\end{itemize}
For each of these uses, the method i.e., AADRNN remains  the same, while the traffic metrics and the decision-making algorithm are adapted to each use.  

\subsection{Botnet Attack Detection with Online Learning}

\begin{figure}
	\centering
	\includegraphics[scale=0.3]{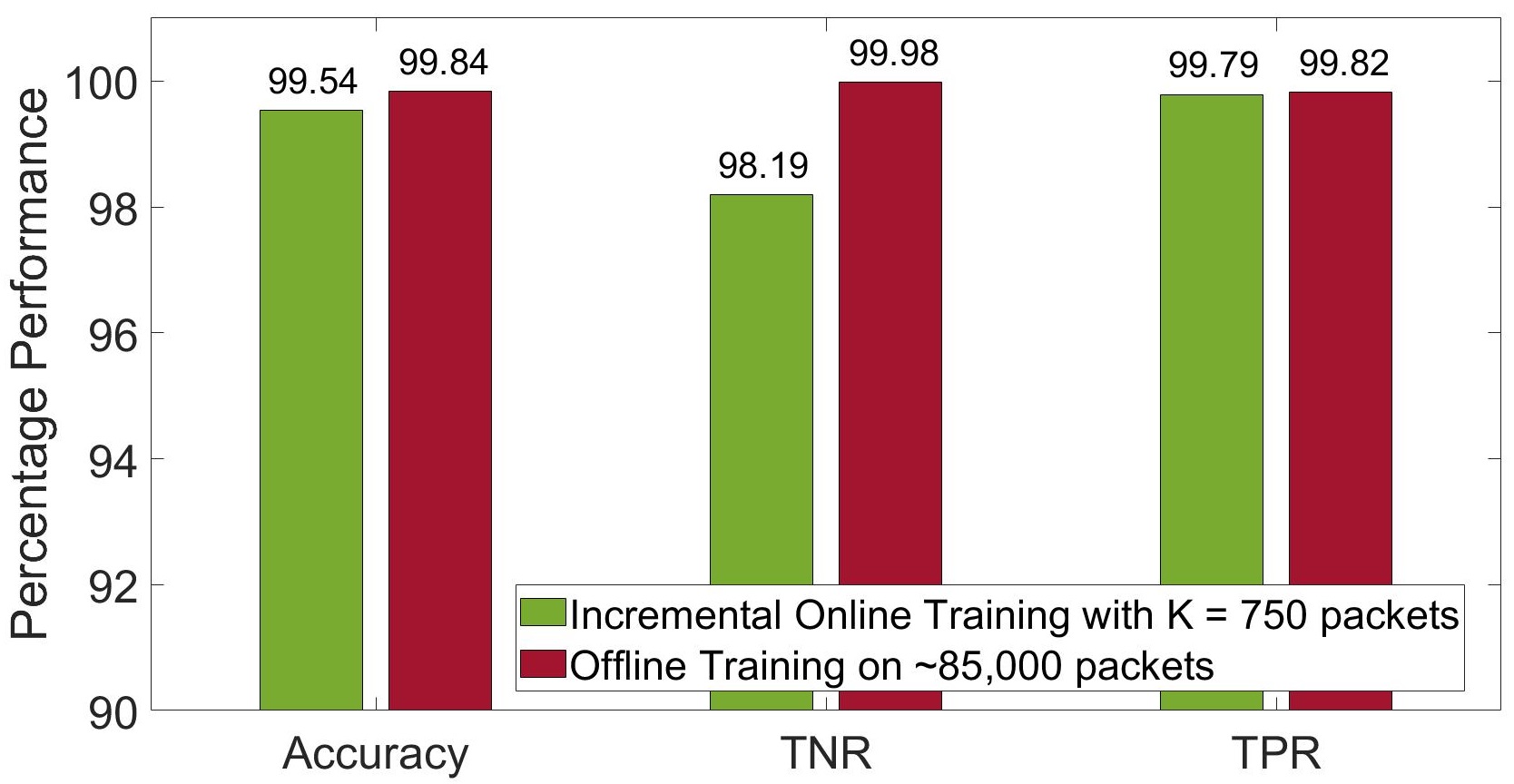}
	\caption{Performance comparison of AADRNN based AD under Incremental Online Training against Offline Training} \label{fig:online}
\end{figure}

The original design of the AD module is targeting the Botnet attack detection. Thus, this module with AADRNN and the proposed traffic metrics was evaluated, as it is, for the detection of the Mirai Botnet attack \cite{nakip_MIRAI, incremental2022} on a publicly available Kitsune dataset \cite{kitsune_dataset, kitsune_paper}. This dataset contains 764,137 transmissions including both normal and attack traffic packets. The evaluation results displayed in Figure~\ref{fig:online} have shown that the incremental online learning property of AD is highly successful with a low false alarm rate as it adapts to the changes in normal traffic over time. 

Table~\ref{table:performance} compares the performance of AADRNN-based AD with that of the state-of-the-art ML methods. The numerical results in this table show the high success of our AD and its superior performance against other ML methods. In detail, we see that our AD based on AADRNN achieves 99.82 \% true positive and 99.98 \% true negative rates.

\begin{table*}[]
	\centering
	\renewcommand{\arraystretch}{1.5}
	\normalsize
	\setlength{\tabcolsep}{6pt} 
	\caption{COMPARISON OF DIFFERENT ML METHODS IN AD MODULE}
	\begin{tabular}{|c|| c || c|c|c|c|}
		\hline
		\begin{tabular}[c]{@{}c@{}}ML \\ Methods\end{tabular} & Accuracy & TPR & FNR & TNR & FPR \\ \hline
		AADRNN & 99.84 & 99.82  & 0.18  & 99.98  & 0.02 \\\hline
		KNN & 99.79 & 99.79 & 0.21 & 99.75 & 0.25\\ \hline
		Lasso  & 99.78 & 99.75 & 0.25 & 99.95 & 0.05\\  \hline
		Simple Thresholding  & 93.18 &  93.09 & 6.94 & 93.63 & 6.37\\ \hline
	\end{tabular}\label{table:performance}
\end{table*}

\subsection{Simultaneous Detection of Different Attack Types}

The AD module was further implemented to detect various types of attacks simultaneously \cite{Gelenbe2022mascots}. To this end, two main revisions have been performed on the design of AD. First, the Metric Extraction module was exchanged with a preprocessing algorithm to apply min-max normalization on any input features provided by the developer (or in a dataset). Then, the decision-making algorithm was enhanced replacing the decision threshold with a statistical whisker which is calculated using the training data. Since extensive threshold selection often provides higher performance for the dataset than the actual test data, data-driven threshold determination, e.g. whisker-valued threshold, eliminates extensive threshold selection.

\begin{figure}[h!]
	\centering
	\includegraphics[scale=0.35]{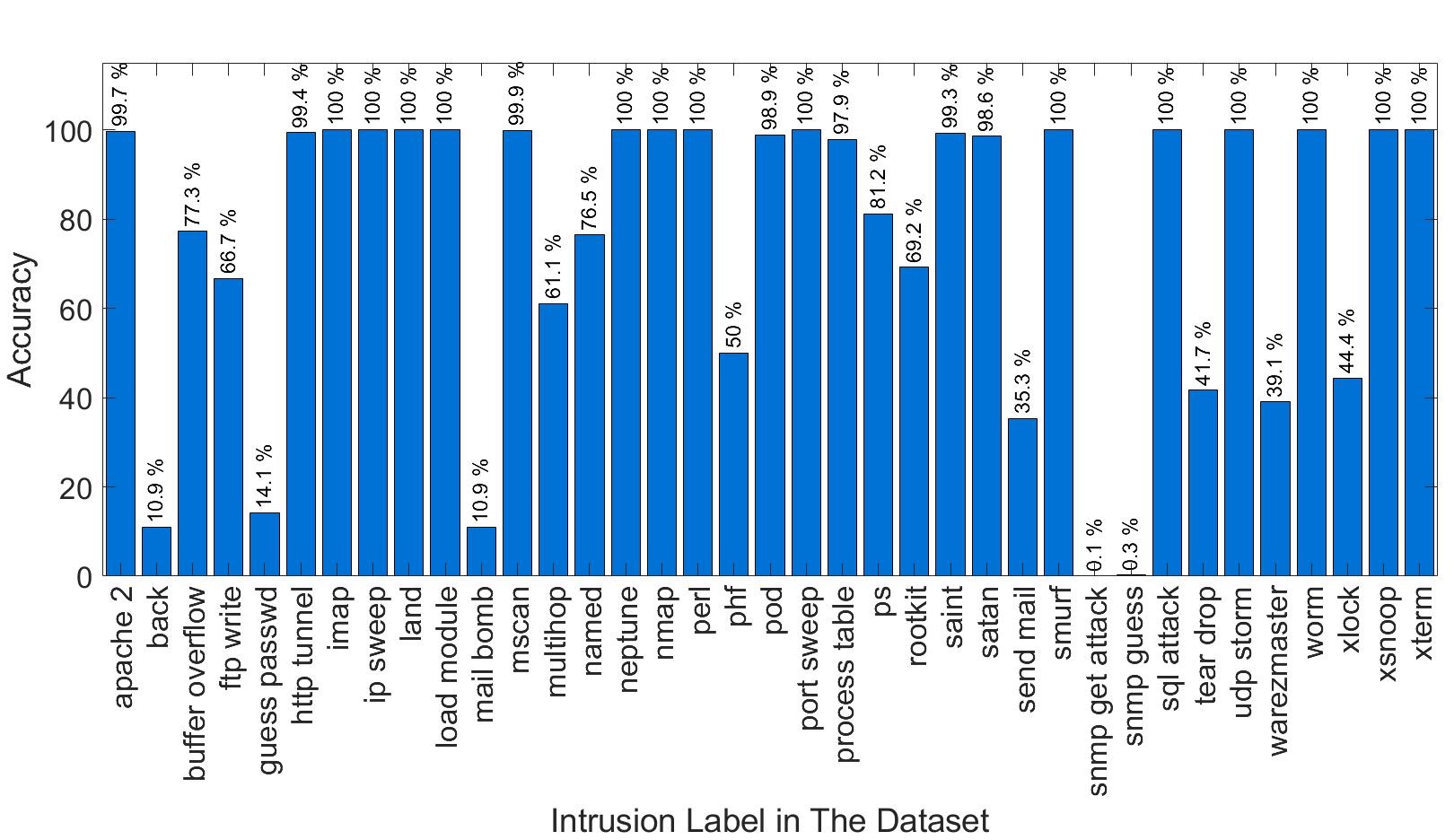}
	\caption{Performance of the AD module for each attack type in the KDD Cup’99 dataset} \label{fig:kdd}
\end{figure}

During the performance evaluation of the AD module to simultaneously detect multiple types of attacks, the KDD Cup’99 data set \cite{kdd} was used. Our AD was trained using only the benign samples in the smaller training set (which is equal to 97,278 samples) and evaluated using all samples in the test set (which is equal to 311,029 samples). As the numerical test results on KDD Cup’99 data set in Figure~\ref{fig:kdd} show, the prediction accuracy of the AD module is above 98 \% for 21 out of 37 attack types.

\subsection{Compromised IoT Device Identification}

The AD methods wre also extended to identify the compromised IoT devices during DDoS attacks, and especially Botnet attacks \cite{access2022}. In this way, the AD module may enable the gateway to blacklist devices affected by malware or take preventive actions against the spread of the attack. To this end, the defined traffic metrics were extended to analyze received and transmitted traffic separately resulting in 6 different metrics. An independent detector was utilized for each existing IP address in the considered IoT network, where each detector was sequentially trained over time. Also, the output of AD was revised to obtain the infection level of the device under consideration.

\begin{figure}[h!]
	\centering
	\includegraphics[scale=0.26]{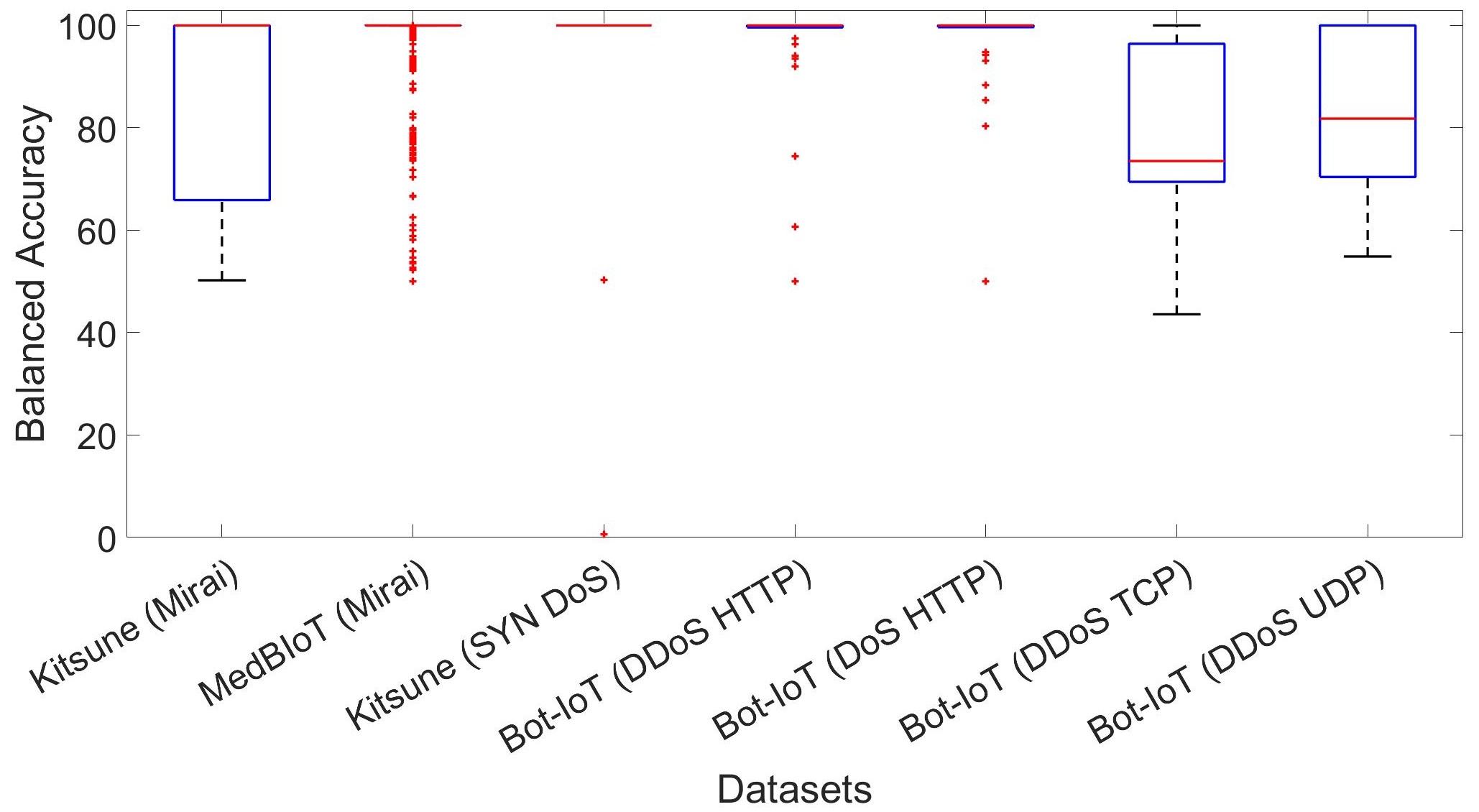}
	\caption{Performance of AD for compromised IoT device identification on Kitsune, MedBIoT, and Bot-IoT data sets} \label{fig:compromised}
\end{figure}

During the evaluation of the performance of the AADRNN based AD, we have used various types of DDoS attacks provided in three different datasets: Kitsune \cite{kitsune_dataset}, MedBIoT \cite{medbiot_dataset}, and Bot-IoT \cite{botiot_dataset}. The results in Figure~\ref{fig:compromised} show that the extended version of the AD module is able to successfully identify compromised devices in the course of various types of DoS and DDoS attacks.

\section{Conclusions}

This paper has reviewed the algorithms developed in the IoTAC project, based on an AADRNN using original network traffic metrics, to detect cyberattacks in real-time with both offline and online learning. Experimental results are summarized, and show that these algorithms are able to {\bf accurately}:
\begin{enumerate}
	\item Detect Botnet attacks with periodically activated incremental learning concurrently with real-time attack detection,
	\item Detect different types of cyberattacks simultaneously using a single AADRNN that is trained offline on normal non-attack traffic, and
	\item Identify compromised IoT devices during an ongoing cyberattack.  
\end{enumerate}
Since the AADRNN is auto-associative, it learns communication patterns for normal (non-attack) network traffic, and differentiates malicious from benign packets (or normal and compromised devices) without prior learning of malicious traffic.  Using the publicly available Kitsune, MedBIoT, BotIoT, and KDD'99 data sets, the three AADRNN based algorithms have been evaluated, and it was observed that all three lead to a high detection accuracy for various types of DoS and DDoS attacks.

Future research will examine methods for post-processing of  the algorithms' output to attempt to further reduce false alarms by an order of magnitude, from the output data of AD algorithms. Another useful direction can be to investigate dynamic system management techniques, such as those investigated in
\cite{SelfAware,Sensors}, 
to mitigate the consequences of cyberattacks using dynamic traffic management for specific IoT systems.

\section*{Acknowledgement}

The authors gratefully acknowledge the support of the European Commission H2020 Program under the IoTAC Research and Innovation Action, under Grant Agreement No. 952684, for the research presented in this paper.

%
%
%
\bibliographystyle{splncs04}
\bibliography{references,security_issues_references,bot_detection_references}

\begin{thebibliography}{10}
\providecommand{\url}[1]{\texttt{#1}}
\providecommand{\urlprefix}{URL }
\providecommand{\doi}[1]{https://doi.org/#1}

\bibitem{kdd}
{KDD Cup 1999 Data},
  \url{http://kdd.ics.uci.edu/databases/kddcup99/kddcup99.html}

\bibitem{kitsune_dataset}
{Kitsune Network Attack Dataset},
  \url{https://www.kaggle.com/ymirsky/network-attack-dataset-kitsune}

\bibitem{hp}
Hp study reveals 70 percent of {Internet of Things} devices vulnerable to
  attack ({Accessed on 25012022}),
  \url{https://www.hp.com/us-en/hp-news/press-release.html\%3Fid=1744676}

\bibitem{Abhishek_gateway}
Abhishek, N.V., Lim, T.J., Sikdar, B., Tandon, A.: An intrusion detection
  system for detecting compromised gateways in clustered iot networks. In: 2018
  IEEE International Workshop Technical Committee on Communications Quality and
  Reliability (CQR). pp.~1--6. IEEE (2018)

\bibitem{Ahmed2019}
Ahmed, Z., Danish, S.M., Qureshi, H.K., Lestas, M.: Protecting {IoT}s from
  {M}irai {B}otnet attacks using blockchains. In: 2019 IEEE 24th International
  Workshop on Computer Aided Modeling and Design of Communication Links and
  Networks (CAMAD). pp.~1--6 (2019). \doi{10.1109/CAMAD.2019.8858484}

\bibitem{Manos}
Antonakakis, M., April, T., Bailey, M., Bernhard, M., Bursztein, E., Cochran,
  J., Durumeric, Z., Halderman, J.A., Invernizzi, L., Kallitsis, M., Kumar, D.,
  Lever, C., Ma, Z., Mason, J., Menscher, D., Seaman, C., Sullivan, N., Thomas,
  K., Zhou, Y.: Understanding the {M}irai {B}otnet. In: Proceedings of the 26th
  USENIX Security Symposium (2017),
  \url{https://www.usenix.org/conference/usenixsecurity17/technical-sessions/presentation/antonakakis}

\bibitem{Banerjee}
Banerjee, M., Samantaray, S.: Network traffic analysis based iot botnet
  detection using honeynet data applying classification techniques.
  International Journal of Computer Science and Information Security (IJCSIS)
  \textbf{17}(8) (2019)

\bibitem{Benzarti}
Benzarti, S., Triki, B., Korbaa, O.: A survey on attacks in {Internet of
  Things} based networks. In: 2017 International conference on engineering \&
  MIS (ICEMIS). pp.~1--7. IEEE (2017)

\bibitem{Biggs}
Biggs, J.: Hackers release source code for a powerful {DD}o{S} app called
  {M}irai. TechCrunch  (October 2018),
  \url{https://techcrunch.com/2016/10/10/hackers-release-source-code-for-a-powerful-ddos-app-called-mirai/}

\bibitem{Brun}
Brun, O., Yin, Y., Gelenbe, E., Kadioglu, Y.M., Augusto-Gonzalez, J., Ramos,
  M.: Deep learning with dense random neural networks for detecting attacks
  against {I}o{T}-connected home environments. In: International ISCIS Security
  Workshop. pp. 79--89. Springer, Cham (2018)

\bibitem{Carl}
Carl, G., Kesidis, G., Brooks, R., Rai, S.: {Denial-of-Service}
  attack-detection techniques. IEEE Internet Computing  \textbf{10}(1),  82--89
  (2006). \doi{10.1109/MIC.2006.5}

\bibitem{Chatterjee_bot}
Chatterjee, M., Namin, A.S., Datta, P.: Evidence fusion for malicious bot
  detection in iot. In: 2018 IEEE International Conference on Big Data (Big
  Data). pp. 4545--4548 (2018). \doi{10.1109/BigData.2018.8621895}

\bibitem{CISA}
CISA: Understanding {Denial-of-Service} attacks,
  \url{https://us-cert.cisa.gov/ncas/tips/ST04-015}

\bibitem{Cisco2020}
Cisco: Cisco Annual Internet Report (2018–2023)  (Mar 2020),
  \url{https://www.cisco.com/c/en/us/solutions/collateral/executive-perspectives/annual-internet-report/white-paper-c11-741490.html}

\bibitem{Douligeris}
Douligeris, C., Mitrokotsa, A.: {DDoS} attacks and defense mechanisms:
  classification and state-of-the-art. Computer networks  \textbf{44}(5),
  643--666 (2004)

\bibitem{Spilios}
Evmorfos, S., Vlachodimitropoulos, G., Bakalos, N., Gelenbe, E.: Neural network
  architectures for the detection of syn flood attacks in {IoT} systems. In:
  Proceedings of the 13th ACM International Conference on PErvasive
  Technologies Related to Assistive Environments. pp.~1--4 (2020)

\bibitem{Sensors}
Fr{\"o}hlich, P., Gelenbe, E., Fio{\l}ka, J., Checinski, J., Nowak, M., Filus,
  Z.: Smart sdn management of fog services to optimize qos and energy. Sensors
  \textbf{21}(9), ~3105 (2021), \url{https://doi.org/10.3390/s21093105}

\bibitem{Frustaci}
Frustaci, M., Pace, P., Aloi, G., Fortino, G.: Evaluating critical security
  issues of the {IoT} world: Present and future challenges. IEEE Internet of
  Things Journal  \textbf{5}(4),  2483--2495 (2018).
  \doi{10.1109/JIOT.2017.2767291}

\bibitem{Deep1}
{Gelenbe}, E., {Yin}, Y.: Deep learning with random neural networks. In: 2016
  International Joint Conference on Neural Networks (IJCNN). pp. 1633--1638
  (2016). \doi{10.1109/IJCNN.2016.7727393}

\bibitem{RNN}
Gelenbe, E.: Random neural networks with negative and positive signals and
  product form solution. Neural Computation  \textbf{1}(4),  502--510 (1989)

\bibitem{SelfAware}
Gelenbe, E., Domanska, J., Frohlich, P., Nowak, M., Nowak, S.: Self-aware
  networks that optimize security, qos and energy. Proceedings of the IEEE
  \textbf{108}(7),  1150--1167 (2020)

\bibitem{Gelenbe2022mascots}
Gelenbe, E., Nak{\i}p: G-networks can detect different types of cyberattacks.
  In: 2022 30th International Symposium on Modeling, Analysis, and Simulation
  of Computer and Telecommunication Systems (MASCOTS). pp.~1--8. IEEE (2022)

\bibitem{access2022}
Gelenbe, E., Nakip, M.: Traffic based sequential learning during botnet attacks
  to identify compromised iot devices. IEEE Access  (2022)

\bibitem{RNN2}
Gelenbe, E., Stafylopatis, A.: Global behavior of homogeneous random neural
  systems. Applied mathematical modelling  \textbf{15}(10),  534--541 (1991)

\bibitem{Deep2}
Gelenbe, E., Yin, Y.: Deep learning with dense random neural networks. In:
  International Conference on Man--Machine Interactions. pp. 3--18. Springer
  (2017)

\bibitem{medbiot_dataset}
Guerra-Manzanares, A., Medina-Galindo, J., Bahsi, H., N{\~o}mm, S.: {MedBIoT}:
  Generation of an {IoT} botnet dataset in a medium-sized iot network. In:
  ICISSP. pp. 207--218 (2020)

\bibitem{Hackett}
Hackett, R.: Why a hacker dumped code behind colossal website-trampling botnet
  (October 2016)

\bibitem{Htwe}
Htwe, C.S., Thant, Y.M., Thwin, M.M.S.: Botnets attack detection using machine
  learning approach for iot environment. In: Journal of Physics: Conference
  Series. vol.~1646, p. 012101. IOP Publishing (2020)

\bibitem{botiot_dataset}
Koroniotis, N., Moustafa, N., Sitnikova, E., Turnbull, B.: Towards the
  development of realistic botnet dataset in the internet of things for network
  forensic analytics: {Bot-IoT} dataset. Future Generation Computer Systems
  \textbf{100},  779--796 (2019).
  \doi{https://doi.org/10.1016/j.future.2019.05.041},
  \url{https://www.sciencedirect.com/science/article/pii/S0167739X18327687}

\bibitem{kumar2019}
Kumar, A., Lim, T.J.: Early detection of {M}irai-like {IoT} {B}ots in
  large-scale networks through sub-sampled packet traffic analysis (2019)

\bibitem{Kumar_MiraiBots}
Kumar, A., Lim, T.J.: Early detection of mirai-like iot bots in large-scale
  networks through sub-sampled packet traffic analysis. In: Future of
  Information and Communication Conference. pp. 847--867. Springer (2019)

\bibitem{Letteri}
Letteri, I., Del~Rosso, M., Caianiello, P., Cassioli, D.: Performance of botnet
  detection by neural networks in software-defined networks. In: ITASEC (2018)

\bibitem{Lin_IoTsurvey}
Lin, J., Yu, W., Zhang, N., Yang, X., Zhang, H., Zhao, W.: A survey on internet
  of things: Architecture, enabling technologies, security and privacy, and
  applications. IEEE Internet of Things Journal  \textbf{4}(5),  1125--1142
  (2017). \doi{10.1109/JIOT.2017.2683200}

\bibitem{Liu_botnet}
Liu, J., Liu, S., Zhang, S.: Detection of iot botnet based on deep learning.
  In: 2019 Chinese Control Conference (CCC). pp. 8381--8385. IEEE (2019)

\bibitem{Margolis}
Margolis, J., Oh, T.T., Jadhav, S., Kim, Y.H., Kim, J.N.: An in-depth analysis
  of the mirai botnet. In: 2017 International Conference on Software Security
  and Assurance (ICSSA). pp. 6--12. IEEE (2017)

\bibitem{McDermott}
McDermott, C.D., Majdani, F., Petrovski, A.V.: Botnet detection in the
  {Internet of Things} using deep learning approaches. In: 2018 international
  joint conference on neural networks (IJCNN). pp.~1--8. IEEE (2018)

\bibitem{kitsune_paper}
Mirsky, Y., Doitshman, T., Elovici, Y., Shabtai, A.: Kitsune: An ensemble of
  autoencoders for online network intrusion detection. In: The Network and
  Distributed System Security Symposium (NDSS) 2018 (2018)

\bibitem{nakip_MIRAI}
Nakip, M., Gelenbe, E.: {MIRAI} botnet attack detection with auto-associative
  dense random neural network. In: IEEE Global Communications Conference
  (Globecom). pp.~1--6 (2021)

\bibitem{incremental2022}
Nakip, M., Gelenbe, E.: Botnet attack detection with incremental online
  learning. In: International ISCIS Security Workshop. pp. 51--60. Springer
  (2022)

\bibitem{Nguyen_federated}
Nguyen, T.D., Marchal, S., Miettinen, M., Fereidooni, H., Asokan, N., Sadeghi,
  A.R.: DÏot: A federated self-learning anomaly detection system for iot. In:
  2019 IEEE 39th International Conference on Distributed Computing Systems
  (ICDCS). pp. 756--767 (2019). \doi{10.1109/ICDCS.2019.00080}

\bibitem{Parra}
Parra, G.D.L.T., Rad, P., Choo, K.K.R., Beebe, N.: Detecting {Internet of
  Things} attacks using distributed deep learning. Journal of Network and
  Computer Applications  \textbf{163},  102662 (2020)

\bibitem{Prokofiev}
Prokofiev, A.O., Smirnova, Y.S., Surov, V.A.: A method to detect internet of
  things botnets. In: 2018 IEEE Conference of Russian Young Researchers in
  Electrical and Electronic Engineering (EIConRus). pp. 105--108. IEEE (2018)

\bibitem{Sinanovic}
Sinanovi{\'c}, H., Mrdovic, S.: Analysis of mirai malicious software. In: 2017
  25th International Conference on Software, Telecommunications and Computer
  Networks (SoftCOM). pp.~1--5. IEEE (2017)

\bibitem{Soe}
Soe, Y.N., Feng, Y., Santosa, P.I., Hartanto, R., Sakurai, K.: Machine
  learning-based iot-botnet attack detection with sequential architecture.
  Sensors  \textbf{20}(16), ~4372 (2020)

\bibitem{Sriram}
Sriram, S., Vinayakumar, R., Alazab, M., Soman, K.: Network flow based iot
  botnet attack detection using deep learning. In: IEEE INFOCOM 2020-IEEE
  Conference on Computer Communications Workshops (INFOCOM WKSHPS). pp.
  189--194. IEEE (2020)

\bibitem{Taneja}
Taneja, M.: An analytics framework to detect compromised iot devices using
  mobility behavior. In: 2013 International Conference on ICT Convergence
  (ICTC). pp. 38--43 (2013). \doi{10.1109/ICTC.2013.6675302}

\bibitem{Tuan}
Tuan, T.A., Long, H.V., Kumar, R., Priyadarshini, I., Son, N.T.K., et~al.:
  Performance evaluation of botnet ddos attack detection using machine
  learning. Evolutionary Intelligence pp. 1--12 (2019)

\bibitem{Tushir_impactsOfMirai}
Tushir, B., Sehgal, H., Nair, R., Dezfouli, B., Liu, Y.: The impact of dos
  attacks onresource-constrained iot devices: A study on the mirai attack.
  arXiv preprint arXiv:2104.09041  (2021)

\bibitem{Tzagkarakis}
Tzagkarakis, C., Petroulakis, N., Ioannidis, S.: Botnet attack detection at the
  iot edge based on sparse representation. In: 2019 Global IoT Summit (GIoTS).
  pp.~1--6. IEEE (2019)

\bibitem{Yang_Securitysurvey}
Yang, Y., Wu, L., Yin, G., Li, L., Zhao, H.: A survey on security and privacy
  issues in internet-of-things. IEEE Internet of Things Journal  \textbf{4}(5),
   1250--1258 (2017). \doi{10.1109/JIOT.2017.2694844}

\end{thebibliography}

\end{document}